# Predicting 3D structure, flexibility and stability of RNA hairpins in monovalent and divalent ion solutions

Ya-Zhou Shi[1], Lei Jin[1], Feng-Hua Wang[2], Xiao-Long Zhu[3], and Zhi-Jie Tan[1*]

[1]*Department of Physics and Key Laboratory of Artificial Micro & Nano-structures of Ministry of Education, School of Physics and Technology, Wuhan University, Wuhan 430072, China*
[2]*Engineering Training Center, Jianghan University, Wuhan 430056, China*
[3]*Department of Physics, School of Physics & Information Engineering, Jianghan University, Wuhan 430056, China*

Running title: A coarse-grained model for RNA structure

## ABSTRACT

A full understanding of RNA-mediated biology would require the knowledge of three-dimensional (3D) structures, structural flexibility and stability of RNAs. To predict RNA 3D structures and stability, we have previously proposed a three-bead coarse-grained predictive model with implicit salt/solvent potentials. In this study, we will further develop the model by improving the implicit-salt electrostatic potential and involving a sequence-dependent coaxial stacking potential to enable the model to simulate RNA 3D structure folding in divalent/monovalent ion solutions. As compared with the experimental data, the present model can predict 3D structures of RNA hairpins with bulge/internal loops (<77nt) from their sequences at the corresponding experimental ion conditions with an overall improved accuracy, and the model also makes reliable predictions for the flexibility of RNA hairpins with bulge loops of different length at extensive divalent/monovalent ion conditions. In addition, the model successfully predicts the stability of RNA hairpins with various loops/stems in divalent/monovalent ion solutions.

[*] To whom correspondence should be addressed. Email: zjtan@whu.edu.cn

# I. INTRODUCTION

In early years, RNA has been considered as an intermediary in transcription and translation (1). However, in the recent two decades, RNAs have been shown to perform other crucial functions such as catalyzing biological reactions and controlling gene expression (2,3). Understanding and utilizing the functions would require the comprehensive knowledge of RNA structures and dynamics (4-7). Although RNA sequences are being discovered rapidly, only limited RNA three-dimensional (3D) structures have been determined through experimental methods such as X-ray crystallography, nuclear magnetic resonance (NMR) spectroscopy and cryo-electron microscopy (4,8). Simultaneously, for high efficiency and low cost, some computational models have been developed for predicting 3D structures or thermodynamics of RNAs (8-18).

Some models based on fragment assembly, sequence alignment and secondary structure are highly successful in predicting 3D structures for even large RNAs (19-39), e.g., MC-Fold/MC-Sym pipeline (30). However, these models are primarily designed to predict folded structures and would not give reliable predictions for the dynamic and thermodynamic properties of RNAs in three dimensions (16-18). Simultaneously, some other models have been developed aiming to predict RNA dynamics and thermodynamics. The Go-like coarse-grained (CG) model of TIS can predict folding thermodynamics for hairpins and pseudoknots (40,41). Another CG model of oxRNA can capture the thermodynamic and mechanical properties of RNA structures with pair-wise interaction potentials (42). But neither of the TIS and oxRNA could give reliable predictions for 3D structures of RNAs from the sequences (16,18). Although the three-bead CG model of iFoldRNA (43) and the six/seven-bead CG model of HiRE-RNA (16,44) can predict 3D structures of small RNAs including pseudoknots, the parameters of the two models may need further validation or adjustment for predicting thermodynamic and dynamic properties of RNAs (18,43,44). Furthermore, since RNAs are highly charged polyanionic polymers, RNA structures can be sensitive to ion conditions and temperature (15,45-53). However, none of the above models could predict 3D structures and thermodynamics of RNAs over a wide range of ion concentrations and temperature from the sequences (16-18).

Very recently, to predict 3D structures and thermal stability of RNAs, we have developed a CG model with three beads placed on the existing atoms P, C4' and N9 for the purine (or N1 for the pyrimidine) (18,54), respectively. Combined with an implicit-salt/solvent force field and the Monte Carlo (MC) simulated annealing algorithm, the model can not only predict native-like 3D structures of small RNAs from their sequences at a high salt (e.g., 1M NaCl), but also give reliable predictions on the stability for RNA hairpins over a wide range of sequences and monovalent ion concentrations as compared with extensive experimental data (54). However, compared with monovalent ions (e.g., $Na^+$), divalent ions such as $Mg^{2+}$ can play a more special role in the stability and dynamics of RNA structures (55-60). For example, $Mg^{2+}$ is about 1000 times more efficient in inducing the tertiary structures folding of *Tetrahymena thermophila* ribozyme (45,49). Although a recent structure-based model with an explicit treatment of $Mg^{2+}$ and an implicit treatment of $K^+$ can well capture the ion atmosphere around RNAs (61-63), there is still lack of a model for predicting 3D structures and stability for RNAs in divalent/monovalent ion solutions from the sequences. In the present work, we will further develop our previous model to enable it to predict the 3D structure and stability of RNAs in the presence of divalent ions.

Additionally, the functions of RNAs may be not only related to the static 3D structures, but also influenced by the flexibility and stability of their structures (4-7,64-66). The flexibility of RNAs is rather important in the recognition with protein and in gene regulation (64-67). For example, the transactivator response element (TAR) for the transactivator (TAT) protein of the human immunodeficiency virus (HIV) can undergo large conformational changes through the small bulge during the binding of TAT proteins (66,67). Due to the polyanionic nature of RNAs, their flexibility would strongly depend on metal ions such as $Mg^{2+}$ and loops (67-69). To examine the effects of metal ions (e.g., $Mg^{2+}$) and loops on RNA flexibility, we will select HIV-1 TAR and HIV-2 TAR variants as two paradigms in the present work. In order to predict the 3D structures and flexibility of RNA hairpins with bulge loops such as HIV TAR variants in divalent/monovalent ion solutions, we will introduce a new implicit electrostatic potential and an indispensable coaxial stacking interaction between two helices at the junction in the present work. With the present model, we will predict the flexibility of HIV-1 TAR and HIV-2 TAR variants in divalent/monovalent ion solutions to understand the effects of salt and bulge loop.

In this work, we will essentially develop the model to simulate RNA folding in divalent/monovalent ion solutions through improving the implicit-salt electrostatic potential and involving a parameterized coaxial stacking potential. Afterwards, we will firstly show the present CG model can predict 3D structures of RNAs with bulge/internal loops at given ionic conditions with higher accuracy. Secondly, the model will be employed to investigate the effects of divalent/monovalent salts and bulge length on the flexibility of HIV TAR variant RNAs. Finally, the model will be used to quantitatively examine the stability of various RNA hairpins in divalent ion solutions. Throughout the paper, all the predictions will be compared with the extensive experimental data.

## II. MODEL AND METHODS

### A. Coarse-grained structural model

Since CG models generally allow considerable extension of the accessible size and time scale in simulations of biological systems (70-74), we have proposed a three-bead CG model for RNAs where three beads stand for phosphate, sugar and base, respectively (54). The backbone phosphate (P) bead and sugar (C) bead are placed respectively at P and C4' atom positions, while the base (N) beads are placed at N9 position for purine or N1 for pyrimidine; see Fig. 1. The P, C and N beads are treated as the spheres with van der Waals radii of 1.9 Å, 1.7 Å and 2.2 Å, respectively (54,75).

## B. Force field

In our CG model, the implicit-solvent/salt force field includes eight energy potentials (54):

$$U = U_b + U_a + U_d + U_{exc} + U_{bp} + U_{bs} + U_{el} + U_{cs}. \tag{1}$$

The function forms for the eight energy potentials are described detailly in Supporting Material, and in the following we only introduce them briefly except for the electrostatic interaction $U_{el}$ and the coaxial stacking interaction $U_{cs}$. The first three terms in Eq. 1 are the bonded potentials for covalent bonds ($U_b$), bond angles ($U_a$), and dihedral angles ($U_d$), respectively. The bonded potentials whose function forms have been shown in Ref. 54, were initially parameterized by the statistical analysis on the available 3D structures of RNA molecules in Protein Data Bank (PDB, http://www.rcsb.org/pdb/home/home.do) (54,75-77). Since lots of native structures in PDB are mostly A-form helix, the statistical parameters from these structures would not be reasonable to describe the nature of RNA free chains (77). Therefore, for bonded potentials, two sets of parameters are calculated for single-strands/loops and stems, named as Para$_{nonhelical}$ and Para$_{helical}$, respectively; see Supporting Material and Ref. 54 for details. The former are used to describe the folding of an RNA from a free chain, and the latter are only used for stems during structure refinement after the folding process. The remaining terms of Eq. 1, namely the nonbonded potentials, describe various pairwise nonbonded interactions. The excluded volume $U_{exc}$ between CG beads is modeled by a purely repulsive Lennard-Jones potential (54,75). $U_{bp}$ in Eq. 1 is employed to capture base-pairing interaction between Watson-Crick (G-C, A-U) and wobble (G-U) base pairs (43,54,78-80). $U_{bs}$ in Eq. 1 is a temperature-dependent base-stacking potential which works between nearest-neighbor base pairs. The strength of $U_{bs}$ was derived from the combined analysis of available sequence-dependent thermodynamic parameters (78-80) and the MC algorithm, and the details are shown in Ref. 54.

The electrostatic interaction $U_{el}$ in Eq. 1, which is a newly refined term for the effect of divalent ions, is taken into account with the combination of the Debye-Hückel (DH) approximation and the concept of counterion condensation (CC) (78,81-83):

$$U_{el} = \sum_{i<j}^{N_P} \frac{(Qe)^2}{4\pi\varepsilon_0\varepsilon(T) r_{ij}} e^{-\frac{r_{ij}}{l_D}}. \tag{2}$$

The summation is over all the phosphate beads and $r_{ij}$ is the distance between two phosphate beads $i$ and $j$. $\varepsilon_0$ is the permittivity of vacuum. $\varepsilon(T)$ is an effective temperature-dependent dielectric constant (40,54,58). $l_D$ is the Debye length of ionic solution. Beyond our previous model (54), the effect of pure divalent ions and the competition between monovalent and divalent ions are also taken into account in the present model to study RNA folding in pure and mixed divalent ion solutions. Based on the CC theory (83), for a pure salt solution containing only one species of salt such as NaCl or MgCl$_2$, the reduced charge fraction $Q$ could be written as $Q = b/(vl_B)$ (40,54), where $v$ is the cation valence. $b$ is the phosphate-phosphate spacing of an RNA and $l_B$ is the Bjerrum length (40,54). For a mixed Na$^+$/Mg$^{2+}$ ion solution, we assume $Q = f_{Na^+} Q_{Na^+} + (1 - f_{Na^+})Q_{Mg^{2+}}$, where $f_{Na^+}$ and $(1 - f_{Na^+})$ represent the contribution fractions from Na$^+$ and Mg$^{2+}$, respectively. $f_{Na^+}$ can be approximately calculated by the empirical formula previously derived from the tightly bound ion (TBI) model which could account for the divalent ion-RNA interactions (15,60,84,85)

$$f_{Na^+} = \frac{[Na^+]}{[Na^+] + x[Mg^{2+}]}. \qquad (3)$$

Here, $x = (8.1 - 64.8/N)(5.2 - \ln[Na^+])$ (15,60). [Na$^+$] and [Mg$^{2+}$] are the corresponding bulk concentrations in molar (M), and $N$ is the chain length.

$U_{cs}$ in Eq. 1 is newly introduced to model the coaxial stacking interaction at RNA junctions (4,86,87), and the coaxial stacking interaction between two discontinuous neighbour helices with interfaced base-pairs $i$-$j$ and $k$-$l$ can be given by (86,87)

$$U_{cs} = \frac{1}{2} \sum_{i-j,k-l}^{N_{cst}} |G_{i-j,k-l}| \{[1 - e^{-a(r_{ik}-r_{cs})}]^2 + [1 - e^{-a(r_{jl}-r_{cs})}]^2 - 2\}, \qquad (4)$$

where $G_{i-j,k-l}$ is the sequence-dependent base-stacking strength. $G_{i-j,k-l}$ is approximately taken as the stacking strength between the corresponding nearest-neighbour base-pairs in an uninterrupted helix (79,86,87). $r_{ik}$ (or $r_{jl}$) is the distance between two interfaced bases $i(j)$ and $k(l)$ of two stems and $a$ represents the extent of distance constraint. $r_{cs}$ is the optimum distance between two coaxially stacked stems. $a$ and $r_{cs}$ are directly obtained from the statistical analysis on the known structures in PDB database (see Fig. S1 in Supporting Material). Here, we only include the cases that there is one base or less in at least one of the single-stranded chains between two neighbour helices (see Fig. S1 in Supporting Material), since the noncanonical base pairs would be generally formed when there are more than one bases in each side between two stems (3-4,86).

The detailed descriptions of the potentials in Eq. 1 and all the parameters for the potentials have been described in Supporting Material; see also Ref. 54 for the building process of force field.

**C. Simulation algorithm**

We use the MC simulated annealing algorithm, which can effectively avoid the trap in local energy minima (34,54,88), to search near-native conformations for an RNA at a given solution condition. Based on a random chain generated for an RNA sequence, the MC simulated annealing algorithm is performed from an initial high temperature to the target temperature (e.g., 298K) at a fixed ion condition. In the folding process, the Para$_{nonhelical}$ of bonded parameters and the efficient pivot moves for RNA chain as well as the standard Metropolis algorithm are used to sample conformations of a free RNA chain (54,75,78). With gradual cooling of the system, the initial 3D near-native structures would be folded at room temperature for RNAs.

After the MC annealing process with the bonded parameters of $Para_{nonhelical}$ for the whole RNA chain, the secondary structure and native-like 3D structures are predicted from a given sequence. To better capture the geometry of helical parts, the further structure refinement is performed for higher accuracy of 3D structures as follows: based on the final 3D structure predicted by the preceding annealing process, another MC simulation (generally $1\times10^6$ steps) is performed at the corresponding ion condition and room temperature, with the bonded parameters of $Para_{helical}$ and $Para_{nonhelical}$ for the base-pairing regions (stems) and loops/single-strands, respectively (54). As a result, an ensemble of refined 3D structures (about eight thousand structures) would be obtained over the last $\sim 8\times10^5$ MC steps. The predicted 3D structures are evaluated by their RMSD's calculated over C beads from the corresponding atoms C4' in the native structure in PDB, though other parameters can also be used to evaluate the predicted structures such as the TM-score and the base interaction network fidelity (89,90). Since the present model generally predicts a series of native-like structures during refinement process, we will use mean RMSD (the averaged value over the whole structure ensemble in refinement process) and minimum RMSD (corresponds to the structure closest to the native one in refinement process) to evaluate the reliability of predictions on 3D structures (29,43,54). Although RNA hairpins are mostly A-form helix, since helical stems may deform from the standard A-form one (48) and the relative orientation between stems at a junction may change (67) upon ionic conditions, we still use the RMSD of the whole RNA including stem and loop to evaluate the performance of 3D structure prediction (89).

**III. RESULTS AND DISCUSSION**

In the following, we will employ the CG model to predict the 3D structures of 32 RNA hairpins most of which are with bulge/internal loops (≤77nt) at the respective experimental ion ($Na^+/Mg^{2+}$) conditions from their sequences. Afterwards, the CG model will be used to predict the flexibility and stability of RNA hairpins at extensive divalent/monovalent ion conditions. Our predictions will be compared with the extensive experiments.

**A. Predicting RNA 3D structures at the respective experimental ion conditions**

Beyond our previous work focused on predicting RNA structures at 1M NaCl (54), we will predict the 3D structure of 32 RNA hairpins most of which are with bulge/internal loops. The 3D structures of these RNAs have been determined by NMR method at certain ion conditions; see Table I. For each RNA, we will make two separate predictions using the present model at the experimental ion condition and its previous version at 1M NaCl (54), respectively. Table I summarizes the major information of the RNA molecules, the corresponding experimental monovalent/divalent ion conditions and the predictions from our model and MC-Fold/MC-Sym pipeline (30).

### 1. *In monovalent solutions*

Firstly, we employ the present model to predict 3D structures for RNA hairpins (with bulge/internal loops) at the corresponding monovalent ionic conditions listed in their PDB files. For example, the structure of stem loop IIa (PDB code: 1u2a) is determined by NMR in the buffer including 10mM $KH_2PO_4$, 50mM KCl, 15mM NaCl and 0.5mM EDTA (91). We predict the 3D structures of the stem loop IIa from its sequence in the solution of 75mM monovalent salt. As shown in Table I, the mean RMSD of the predicted 3D structures is ~2.1 Å, which is obviously smaller than that (~2.6 Å) of the structures predicted by the previous version of our model at 1 M NaCl. For the 28 tested RNA hairpins in the respective monovalent salt solutions, the overall mean RMSD between the structures predicted by the present model at the respective experimental monovalent ionic conditions and the experimental structures is 3.37 Å, a smaller value than that (3.66 Å) predicted by the previous version of the model at 1 M NaCl (54); see Table I.

### 2. *In divalent solutions*

In addition, one important feature of the present model is the combination between the CC theory and the results from the TBI model, and the present model can be employed to simulate RNA folding in mixed monovalent/divalent ion solutions. Four RNA hairpins with bulge/internal loops (PDB code: 1yn2, 2l5z, 2aht and 1p5o) have been determined by NMR in solutions containing $Mg^{2+}$ and the corresponding monovalent/divalent ion conditions are listed in Table I. The mean RMSDs for these RNAs predicted by the present model in the corresponding experimental mixed ion solutions are 4.1 Å, 4.0 Å, 3.2 Å and 11.0 Å, respectively, which are apparently smaller than the values (4.3 Å, 4.6 Å, 3.9 Å and 13.0 Å) predicted by our previous model regardless of salt effect and the coaxial stacking interactions (54).

### 3. *Salt vs coaxial stacking*

As shown in Table I, for the 32 hairpins, the present model can predict a visibly lower overall mean RMSD than our previous model (3.64 Å vs 4.02 Å; the *P*-value of <0.01 from the two-tailed Student's *t*-test (12)), which suggests that the involvement of monovalent/divalent salt and coaxial stacking in the present model is effective for predicting RNA 3D structures in ion solutions. To clarify the contributions of the two improvements, we further perform two additional predictions for each hairpin using the present model with coaxial stacking at 1M NaCl and the present model without coaxial stacking at the experimental ion conditions, respectively. We find that the two improvements would both make positive contributions to the overall improved predictions and the involvement of salt effect has stronger contribution than that of coaxial stacking, as shown in Fig. S2 in Supporting Material. Due to the high charge density of RNA backbone, the native-like structures of hairpins could be slightly stretched at low salt (48), and the difference between conformations at low and high salts should not be ignored (see Fig. S2). Although the coaxial stacking potential would only have slight effect on hairpins with internal/bulge loops whose bases are stacking into neighbor helices (e.g., 1jur and 1lc6), it would be indispensable for the formation of coaxially stacked states which are common in RNAs with bulges (e.g., 2kuu and 1p5o), and in the HIV-1 and HIV-2 TAR variants; see the subsection of "flexibility of RNAs with bulges".

*4. Comparisons with MC-Fold/MC-Sym pipeline*

The MC-Fold/MC-Sym pipeline is a web server (http://www.major.iric.ca/MC-Pipeline/) for RNA secondary and tertiary structure prediction (30). To test the present model, we also make comparisons with the MC-Fold/MC-Sym pipeline. The RMSDs of the top 1 structures predicted by the pipeline online server (option: return the best 100 secondary structures and model_limit = 1000 or time_limit = 12 h) are calculated over C4' atoms from the corresponding atoms in the experimental structures. As shown in Table I and Fig. 2, the overall predictions of 32 structures from the present model (overall mean RMSD = 3.64Å) appear visibly better than those from the MC-Fold/MC-Sym pipeline (overall mean RMSD = 4.12Å), which is also suggested by the *P*-value of <0.01 from the two-tailed Student's *t*-test (12). Table I and Fig. 2 also show that the present model can give reliable prediction for relatively large RNAs (>45nt (54)). For the 48-nt and 68-nt RNAs (PDB codes: 2kuu and 2mqt), the present model gives much better predictions than those predicted by the MC-Fold/MC-Sym pipeline, while for the 77-nt RNA (PDB code: 1p5o), the two models give the similar predictions (see Table I and Fig. 2). This may be attributed to that the present model still ignores the possible noncanonical base pairs, which are abundant in the 77-nt RNA.

**B. Flexibility of RNAs with bulges**

RNAs are highly flexible biomolecules that can undergo dramatic conformational changes to fulfil their diverse functions, e.g., the structural reorganization of riboswitches or the hammerhead ribozyme (2,3). Generally, large conformational transitions of RNA structures are induced by the binding of ions, proteins or ligands (45-47,65-67). For example, HIV TAR RNAs, which can bind TAT protein through conformational changes in viral replication, have been the important paradigms for studying RNA dynamics (66,67). In the following, we will employ the present model to study the flexibility of HIV-1/HIV-2 TAR variants at different ion conditions and make comparisons with the experimental data. The sequences of the HIV TAR variants and their secondary structures predicted by the present model are shown in Fig. 3. HIV-1 TAR variant with a 3-nt bulge is used to examine the conformational changes at different ion conditions with the present model, and HIV-2 TAR variants with different length of polyU (polyA) bulges are used to study the conformational changes induced by various bulges. Our predictions are compared with the existing experiments for HIV-1

TAR variant at extensive $Na^+$/$Mg^{2+}$ concentrations and for HIV-2 TAR variant with various length bulges at 5mM [$Na^+$] with/without 2mM [$Mg^{2+}$], respectively (67,68).

*1. Bending versus monovalent/divalent salts*

Recent studies have shown that RNA flexibility is strongly coupled to its ion condition (45-47,67-69,75,92). As Casiano-Negroni et al have experimentally measured, the bending angle at the junction of bulge is strongly dependent on salt (67). Here, we use the present model to evaluate the bending at the bulge junction of the HIV-1 TAR variant (see Fig. 3) over a broad range of [$Na^+$] as well as [$Mg^{2+}$]. As shown in Figs. 4a and 4b, the inter-helical bend angle at the junction of bulge decreases apparently with the increase of [$Na^+$] (and [$Mg^{2+}$]), and our predictions agree well with the corresponding experimental data (67). Such decrease of bending angle with the increase of [$Na^+$]/[$Mg^{2+}$] is understandable. It comes from the competition between electrostatic repulsion which plays a dominate role at low salt and coaxial stacking interaction at the bulge which plays a dominate role at high salt. The comparison between Fig. 4a and Fig. 4b shows that $Na^+$ and $Mg^{2+}$ induce a similar structural transition from a bent state at low ion concentrations to a coaxially stacked state at high salt, while $Mg^{2+}$ is much more efficient in inducing such structure transition due to the higher ionic charge (55-58).

To further clarify the effect of coaxial stacking, we use the present model without coaxial stacking to predict the 3D structures for HIV-1 TAR variant at extensive $Na^+$/$Mg^{2+}$ concentrations. The comparison of inter-helical bend angles predicted by the present model with and without coaxial stacking indicates that the involvement of coaxial stacking can effectively capture the coaxially stacked conformations closer to the experiments especially at high salt (67), as shown in Fig. S3 in Supporting Material.

*2. Bending versus bulge length*

Zacharias and Hagerman have experimentally measured the bending angle for long RNA helices induced by bulges of various length ($n$=1 to 6) and base composition ($A_n$ and $U_n$ series) (68). For simplicity, we predict the 3D structures of HIV-2 TAR variant (see Fig. 3) with different bulge length at 5mM $NaPO_4$ in the absence/presence of 2mM $Mg^{2+}$ to study the bulge-induced bending of RNAs. As shown in Figs. 5a and 5b, the bend angle at the bulge increases with the increase of the number of nucleotides in the bulge, which is in good accordance with the experimental data for the bulge of $U_n$ (68). As shown in Fig. 5a, the severe inter-helical bending and its sharp increase for longer bulge come from the more extended structures with larger end-to-end distance for longer bulges and the strong electrostatic repulsion between helices which can offset the interhelical stacking interactions at low salt concentrations. Higher salt (2mM $Mg^{2+}$) would reduce the electrostatic repulsion more strongly, and thus would promote the interhelical stacking and reduce the bending angles; see Fig. 5b. However, as shown in Figs. 5a and 5b, the predictions on bending angle are apparently smaller than the experimental data for HIV-2 TAR variant with $A_n$ bulge. This is because polyU behaves like a random coil, while polyA would exhibit strong intra-chain self-stacking (69) which is not accounted for in the present model. Such self-stacking would enhance the rigidity of single-stranded chain and consequently cause the large bending angle at the bulge.

Recently, Mustoe et al have also developed a CG model of TOPRNA, which treats RNAs as collections of semirigid helices linked by freely rotatable single strands (74). The model could nearly reproduce experimental bending angles of HIV-2 TAR with more than 2 nt bulges at low salt concentrations, while it did not give reliable predictions for HIV-2 TAR with polyU bulge at high salt as well as HIV-2 TAR with 1 nt bulge, possibly due to the ignorance of salt effect/coaxial stacking and the overestimate of bulge rigidity for polyU (74).

### 3. *Charactering global structural fluctuation*

The global size of an RNA can be characterized by its radius of gyration $R_g$, the fluctuation of which can properly reflect the RNA structural flexibility (64,93). Based on the conformational ensemble of an RNA, we have calculated the variance $\sigma_R^2$ of $R_g$ by $\sigma_R^2 = \overline{(R_g - \overline{R_g})^2}$; see Fig. S4 and Fig. S5 in Supporting Material, as well as the mean RMSD from the time-averaged reference structure. As shown in Figs. 4c and 4d, $\sigma_R^2$ and RMSD of the HIV-1 TAR variant increase with the increase of salt concentration, and such increase becomes saturated at high salts. This is because that the higher salt can reduce the electrostatic repulsion in the RNA and would favor the conformational fluctuation. At high salts, the coaxial stacking between the two stems can be formed and consequently $\sigma_R^2$ (and RMSD) becomes saturated. Furthermore, as shown in Figs. 5c and 5d, $\sigma_R^2$ (and RMSD) of the HIV-2 TAR variant increases for longer bulge length in 5mM $NaPO_4$ solution in the absence/presence of $Mg^{2+}$. This is because RNA single-stranded loops are distinctively more flexible than duplexes (69,75,93).

### 4. *Charactering local structural fluctuation*

Furthermore, we have calculated the root-mean-square-fluctuation (RMSF) of backbone beads to analyze the local flexibility along an RNA chain (64). The RMSF for the *i*-th C-bead is calculated as:

$$RMSF_i = \sqrt{\left\langle \left(r_i(t) - r_i^0\right)^2 \right\rangle_t}, \qquad (5)$$

where $\langle \cdots \rangle_t$ means the average over time *t*. $r_i(t)$ is the position of the *i*-th C-bead at time *t*, and $r_i^0$ is the time-averaged reference position of the *i*-th C-bead.

Fig. 6 shows the RMSF of each C-bead of HIV-1 TAR variant at 1M NaCl. The nucleotides near the 5' and 3' ends exhibit stronger conformational fluctuations, since the terminal base pairs have less spatial constraints (93). As shown in Fig. 6 and Fig. 7, the bulge and hairpin loops are more flexible than the stems, which comes from the significantly higher flexibility of unpaired single-stranded chains, since the previous theoretical and experimental studies have suggested that the persistence length of stem is ~60 times higher than that of single-strand chain (64-69,75,93). When the length of bulges increases from 1 to 8, the fluctuations of nucleotides especially at the bulge are apparently enhanced; see Fig. 7. This is because longer bulge has higher flexibility and naturally causes the higher flexibility of the whole RNA. Additionally, Fig. 7 shows that the addition of $Mg^{2+}$ could increase the local flexibility of RNAs, especially for HIV-2 TAR variant with longer bulge. For the RNAs with small bulges, the salt effect on local flexibility would not be very obvious; see also Fig. S6. It is reasonable since the addition of $Mg^{2+}$ would bring the strong electrostatic screening and consequently increase the flexibility of RNAs, especially at bulges. But for RNAs with small bulges (≤ ~3nt), the coaxial stacking would be formed between two stems and the addition of $Mg^{2+}$ would only have slight effect on their flexibility.

**C. RNA hairpin stability in divalent ion solutions**

RNA folded structure is stabilized by the interplay of diverse interactions such as base pairing/stacking and electrostatic interactions. RNA stability at high salt concentrations (e.g., 1M NaCl) can be predicted with a relatively simple nearest-neighbor model (79,80). However, the nearest-neighbor model cannot predict the 3D structure of RNAs at an arbitrary temperature, and cannot predict RNA stability at ionic conditions departing from 1M NaCl. However, due to the polyanionic nature, RNA stability is very sensitive to the ionic condition (55-58,94-99), and $Mg^{2+}$ ions are particularly efficient in stabilizing RNA tertiary structure (55-57). To address the effect of $Mg^{2+}$ in RNA stability, we will employ the present model to study the stability of various RNA hairpins in divalent and mixed divalent/monovalent ion solutions. Generally, a hairpin is either in folded state at low *T*, or in unfolded state at high *T*, or in bistability at middle *T* around the melting temperature $T_m$ (50-52,54). Based on the equilibrium value of the number of base pairs at each temperature *T*, to obtain the $T_m$ of a hairpin, the fraction of denatured base pairs *f(T)* can be calculated and fitted to a two-state model (54,79),

$$f(T) = 1 - \frac{1}{1 + e^{(T-T_m)/dT}}, \tag{6}$$

where *dT* is an adjustable parameter (54).

*1. In pure divalent solutions*

Recently, the loop-size dependence of the stability of an RNA hairpin (denoted as R0) has been experimentally investigated in 2.5mM [$Mg^{2+}$] (94). Fig. 8a shows the sequences of hairpins R0 as well as the secondary structures predicted by the present model. Fig. 8a also shows the predicted $T_m$ for R0 in 2.5mM [$Mg^{2+}$] as a function of the loop size (m=4~34 nt), which is in good accordance with experimental data (94). Due to the larger conformational entropy for longer loop, the hairpin stability would decrease when the hairpin loop becomes longer. In addition, we have studied the stability of hairpins with the same loop while with different stems in 0.7mM $Mg^{2+}$ solutions. Hairpins R1~R4 are four similar RNAs, whose stems are slightly different in length or sequence; see Fig. 8b. As shown in Fig. 8b, $T_m$'s of four hairpins predicted by the present model at 0.7mM $MgCl_2$ are in good agreement with the experimental data (95). The addition of G-C or C-G base-pair (from R1 to R2 and then to R3) can dramatically stabilize the RNA hairpins due to the strong base pairing/stacking interactions. The difference between $T_m$'s of R3 and R4 indicates that RNA stability is sensitive to sequence-dependent base pairing/stacking. The good agreement between our predictions and the experiments (94,95) suggests that the present model can well describe the folding stability of small RNAs in pure $Mg^{2+}$ solutions.

## *2. In mixed divalent/monovalent solutions*

With the present model, we have also examined the stability of RNA hairpins R5 and R6 in mixed $K^+$/$Mg^{2+}$ solutions, and the secondary structures for R5 and R6 predicted by the present model are shown in Figs. 8c and 8d. As shown in Figs. 8c and 8d for $T_m$'s of R5 and R6 at a fixed [$K^+$] (100mM), in addition to the general trend of increased stability for higher [$Mg^{2+}$], the competition between $K^+$ and $Mg^{2+}$ also leads to the following behavior of RNA stability. At low [$Mg^{2+}$] (≤0.1mM), the stability of RNAs is dominated by (~100mM) $K^+$ and the $T_m$'s are close to that at pure corresponding [$K^+$]. As [$Mg^{2+}$] is increased, $Mg^{2+}$ ions begin to play a role and the stability of RNAs begins to increase markedly due to the efficient role of $Mg^{2+}$ in stabilizing RNAs. At very high [$Mg^{2+}$], $Mg^{2+}$ would become dominated and $T_m$ becomes saturated (60). As shown in Figs. 8c and 8d, the agreements with experimental data (98,99) indicate that the combination of CC theory and the results from the TBI model could give good description for the competition between monovalent and divalent ions in stabilizing RNA hairpins, and the present model can give good predictions for the stability of RNA hairpins in mixed divalent/monovalent solutions.

## IV. CONCLUSIONS

In this work, we have developed our CG model to predict 3D structures and structural properties of RNAs with bulge/internal loops in the presence of divalent and monovalent ions. The major extensions of our CG model include the improvement of the electrostatic potential to implicitly consider the effect of divalent ions and the inclusion of the coaxial stacking at two-way junction. The improved CG model has been employed to examine the effects of divalent/monovalent salt on the 3D structures, flexibility and stability of RNA hairpins with two-way junction.

Firstly, we have employed the present model to predict 3D structures for 32 RNAs (≤ 77nt) at their respective monovalent/divalent salt conditions in which the RNA structures have been experimentally determined by NMR method, and the overall mean RMSD of 3.64 Å between the predictions and the experimental structures is visibly smaller than those from predictions by the MC-Fold/MC-Sym pipeline and by the previous version of our model at 1M NaCl. Secondly, we have studied the flexibility of RNA hairpins with varying bulge loops at extensive [$Na^+$]'s and [$Mg^{2+}$]'s, and the predicted bending angles at bulge for HIV-1 and HIV-2 TAR variants are in good agreement with the available experimental data for different salt conditions as well as the different lengths of bulge loops. Thirdly, we have predicted the stability for RNA hairpins in divalent and mixed divalent/monovalent ion solutions, and the predictions agree well with the experimental data. Therefore, the present model can provide the ensemble of probable 3D structures at extensive divalent/monovalent ion conditions and can make reliable predictions on the structural properties such as flexibility and stability for small RNAs.

Despite of the extensive agreement between our predictions and the experiments, the present model still involves some approximations and simplifications. First, in the present model, the effect of divalent and monovalent salts is implicitly accounted for by the combination of CC theory and TBI model. The good agreement with experimental data suggests that the combination of CC theory and TBI model can well capture the efficient role of divalent ions over monovalent ions, though the more extensive experimental validation for larger RNAs is still required. Also, the present model ignores the effect of specific ion binding, which might become important for large RNAs with complex structures (55-57). Of course, the more accurate treatment on salt is to explicitly consider the metal ions (75,97), which would bring huge computation cost. Very recently, Hayes et al have proposed a generalized Manning counterion condensation model in an alternative way to reproduce the ion atmosphere around RNAs through the explicit representation of $Mg^{2+}$ and implicit treatment of $K^+$ (61-63). Second, the present CG model only considers the canonical Watson-Crick (C-G, A-U) and wobble G-U base pairing, and ignores the possible noncanonical base pairs due to the lack of the experimental thermodynamic parameters (79,80). The noncanonical base pairing can be involved in the present model with the corresponding thermodynamic parameters, which would further improve the accuracy of structure prediction for RNAs with loops (29,30). Third, although our model can predict 3D structures for RNAs beyond hairpins, e.g., small pseudoknots (Ref. 54), it is still with challenge for the model at the present version to accurately and efficiently predict 3D structures of large RNAs with complex structures. Nevertheless, we are currently extending the model to predict 3D structures and stability for extensive RNA pseudoknots, and complex structures, while the 3D structure prediction for large RNAs from sequences may possibly require certain experimental constraints (34,39). Finally, the 3D structure predicted by the present model is at coarse-grained level, and consequently it is still required to develop the present model to reconstruct the all-atomistic structures based on the CG predictions. Nevertheless, the present model can be a reliable predictive model for 3D structure ensemble of small RNAs in divalent/monovalent solutions and at arbitrary temperatures.

## SUPPORTING MATERIAL

More details of force-field and six figures are available at xxxx.

## AUTHOR CONTRIBUTIONS

Z.J.T. and Y.Z.S. designed the research; Y.Z.S., L.J., and F.H.W. performed the research; Z.J.T., Y.Z.S., and X.L.Z. analyzed data; and Y.Z.S. and Z.J.T. wrote the article.

## ACKNOWLEDGEMENTS

We are grateful to Profs. Shi-Jie Chen (Univ Missouri), Yang Zhang (Univ Michigan) and Wenbing Zhang (Wuhan Univ) for valuable discussions. This work was supported by the National Key Scientific Program (973)-Nanoscience and Nanotechnology (No. 2011CB933600), the National Science Foundation of China grants (11175132, 11374234 and 11575128), and the Program for New Century Excellent Talents (Grant No. NCET 08-0408).

## REFERENCES

1. Crick, F. 1970. Central dogma of molecular biology. Nature. 227:561-563.
2. Doherty, E. A., and J. A. Doudna. 2001. Ribozyme structures and mechanisms. Annu. Rev. Biophys. Biomol. Struct. 30:457-475.
3. Edwards, T. E., D. J. Klein, and A. R. Ferre-d'Amare. 2007. Riboswitches: small-molecule recognition by gene regulatory RNAs. Curr. Opin. Chem. Biol. 17:273-279.
4. Tinoco, I., Jr., and C. Bustamante. 1999. How RNA folds. J. Mol. Biol. 293:271-281.
5. Hall, K. B., 2012. Spectroscopic probes of RNA structures and dynamics. Methods Mol. Biol. 875:67-84.
6. Zhang, W., and S. J. Chen. 2002. RNA hairpin-folding kinetics. Proc. Natl. Acad. Sci. USA 99:1931-1936.
7. Gong, S., Y. Wang, and W. Zhang. 2015. Kinetic regulation mechanism of pbuE riboswitch. J. Chem. Phys. 142:015103.
8. Sim, A. Y., P. Minary, and M. Levitt. 2012. Modeling nucleic acids. Curr. Opin. Struct. Biol. 22:273-278.
9. Rother, K., M. Rother, M. Boniecki, T. Puton, and J. M. Bujnicki. 2011. RNA and protein 3D structure modeling: similarities and differences. J. Mol. Model 17:2325-2336.
10. Laing, C., and T. Schlick. 2011. Computational approaches to RNA structure prediction, analysis, and design. Curr. Opin. Struct. Biol. 21:306-318.
11. Cruz, J. A., M. F. Blanchet, M. Boniecki, J. M. Bujnicki, S. J. Chen, S. Cao, R. Das, F. Ding, N. V. Dokholyan, S. C. Flores, L. Huang, C. A. Lavender, V. Lisi, F. Major, K. Mikolajczak, D. J. Patel, A. Philips, T. Puton, J.

Santalucia, F. Sijenyi, T. Hermann, K. Rother, M. Rother, A. Serganov, M. Skorupski, T. Soltysinski, P. Sripakdeevong, I. Tuszynska, K. M. Weeks, C. Waldsich, M. Wildauer, N. B. Leontis, and E. Westhof, 2012. RNA-Puzzles: A CASP-like evalution of RNA three-dimensional structure prediction. RNA. 18:610-625.

12. Hajdin, C. E., F. Ding, N. V. Dokholyan, and K. M. Weeks. 2010. On the significance of an RNA tertiary structure prediction. RNA. 16:1340-1349.

13. Shapiro, B. A., Y. G. Yingling, W. Kasprzak, and E. Bindewald. 2007. Bridging the gap in RNA structure prediction. Curr. Opin. Struct. Biol. 17:157-165.

14. Zhang, Y. 2008. Progress and challenges in protein structure prediction. Curr. Opin. Struct. Biol. 18:342-348.

15. Tan, Z. J., W. Zhang, Y. Z. Shi, and F. H. Wang. 2015. RNA folding: structure prediction, folding kinetics and ion electrostatics. Adv. Exp. Med. Biol. 827:143-183.

16. Cragnolini, T., P. Derreumaux, and S. Pasquali. 2015. Ab initio RNA folding. J. Phys. Condens. Mat. 27:233102.

17. Bailor, M. H., A. M. Mustoe, C. L. Brooks, and H. M. Al-Hashimi. 2011. Topological constraints: using RNA secondary structure to model 3D conformation, folding pathways, and dynamic adaptation. Curr. Opin. Struct. Biol. 21:296–305.

18. Shi, Y. Z., Y. Y. Wu, F. H. Wang, and Z. J. Tan. 2014. RNA structure prediction: progress and perspective. Chin. Phys. B 23:078701.

19. Massire, C., and E. Westhof. 1998. MANIP: an interactive tool for modelling RNA. J. Mol. Graph. Model. 16:197-205.

20. Jossinet, F., T. E. Ludwig, and E. Westhof. 2010. Assemble: an interactive graphical tool to analyze and build RNA architectures at 2D and 3D levels. Bioinformatics 26:2057-2059.

21. Martinez, H. M., J. V. Maizel, and B. A. Shapiro. 2008. RNA 2D3D: a program for generating, viewing, and comparing 3-dimensional models of RNA. J. Biomol. Struct. Dyn. 25:669-683.

22. Rother, M., K. Rother, T. Puton, and J. M. Bujnicki. 2011. ModeRNA: A tool for comparative modeling of RNA 3D structure. Nucleic Acids Res. 39:4007-4022.

23. Flores, S. C., and R. B. Altman. 2010. Turning limited experimental information into 3D models of RNA. RNA. 16:1769-1778.

24. Popenda, M., M. Szachniuk, M. Antczak, K. J. Purzycka, P. Lukasiak, N. Bartol, J. Blazewicz, and R. W. Adamiak. 2012. Automated 3D structure composition for large RNAs. Nucleic Acids Res. 40:e112.

25. Zhao, Y., Z. Gong, and Y. Xiao, 2011. Improvements of the hierarchical approach for predicting RNA tertiary structure. J. Biomol. Struct. Dyn. 28:815-826.

26. Zhao, Y., Y. Huang, Z. Gong, Y. Wang, J. Man, and Y. Xiao. 2012. Automated and fast building of three-dimensional RNA structures. Sci. Rep. 2:734.

27. Huang, Y., S. Liu, D. Guo, L. Li, and Y. Xiao. 2013. A novel protocol for three-dimensional structure prediction of RNA-protein complexes. Sci. Rep. 3:1887.

28. Wang, J., Y. Zhao, C. Zhu, and Y. Xiao. 2015. 3dRNAscore: a distance and torsion angle dependent evalution function of 3D RNA structure. Nucleic Acids Res. 43:e63.

29. Das, R., and D. Baker. 2007. Automated de novo prediction of native-like RNA tertiary structures. Proc. Natl. Acad. Sci. USA 104:14664-14669.

30. Parisien, M., and F. Major. 2008. The MC-Fold and MC-Sym pipeline infers RNA structure from sequence

data. Nature. 452:51-55.

31. Bida, J.P., and L. J. Maher. 2012. Improved prediction of RNA tertiary structure with insights into native state dynamics. RNA. 18:385-393.

32. Zhang, J., Y. Bian, H. Lin, and W. Wang. 2012. RNA fragment modeling with a nucleobase discrete-state model. Phys. Rev. E 85:021909.

33. Zhang, J., J. Dundas, M. Lin, M. Chen, W. Wang, and J. Liang. 2009. Prediction of geometrically feasible three-dimensional structures of pseudoknotted RNA through free energy estimation. RNA. 15:2248-2263.

34. Seetin, M. J., and D. H. Mathews. 2011. Automated RNA tertiary structure prediction from secondary structure and low-resolution restraints. J. Comput. Chem. 32:2232-2244.

35. Tan, R. K. Z., A. S. Petrov, and S. C. Harvey. 2006. YUP: A molecular simulation program for coarse-grained and multiscaled models. J. Chem. Theory Comput. 2:529-540.

36. Jonikas, M. A., R. J. Radmer, A. Laederach, R. Das, S. Pearlman, D. Herschlag, and R. B. Altman. 2009. Coarse-grained modeling of large RNA molecules with knowledge-based potentials and structural filters. RNA. 15:189-199.

37. Cao, S., and S. J. Chen. 2011. Physics-based de novo prediction of RNA 3D structures. J. Phys. Chem. B 115:4216-4226.

38. Xu, X., P. Zhao, and S. J. Chen. 2014. Vfold: A web server for RNA structure and folding thermodynamics prediction. PLoS One 9:e107504.

39. Xia, Z., D. R. Bell, Y. Shi, and P. Ren. 2013. RNA 3D structure prediction by using a coarse-grained model and experimental data. J. Phys. Chem. B 117:3135-3144.

40. Denesyuk, N., and D. Thirumalai. 2013. Coarse-grained model for predicting RNA folding thermodynamics. J. Phys. Chem. B 117:4901–4911.

41. Hyeon, C., and D. Thirumalai. 2005. Mechanical unfolding of RNA hairpins. Proc. Natl. Acad. Sci. USA 102:6789-6794.

42. Sulc, P., F. Romano, T. E. Ouldridge, J. P. K. Doye, and A. A. Louis. 2014. A nucleotide-level coarse-grained model of RNA. J. Chem. Phys. 140:235102.

43. Ding, F., S. Sharma, P. Chalasani, V. V. Demidov, N. E. Broude, and N. V. Dokholyan. 2008. Ab initio RNA folding by discrete molecular dynamics: from structure prediction to folding mechanisms. RNA. 14:1164-1173.

44. Pasquali, S., and P. Derreumaux. 2010. HiRE: A high resolution coarse-grained energy model for RNA. J. Phys. Chem. B 114:11957-11966.

45. Woodson, S. A. 2005. Metal ions and RNA folding: a highly charged topic with a dynamic future. Curr. Opin. Struct. Biol. 9:104-109.

46. Chen, S. J. 2008. RNA folding: conformational statistics, folding kinetics, and ion electrostatics. Annu. Rev. Biophys. 37:197-214.

47. Lipfert, J., S. Doniach, R. Das, and D. Herschlag. 2014. Understanding nucleic acid-ion interactions. Annu. Rev. Biochem. 83:813-841.

48. Manning, G. S. 2014. The response of DNA length and twist to changes in ionic strength. Biopolymers. 103:223-226.

49. Takamoto, K., Q. He, S. Morris, M. R. Chance, and M. Brenowitz. 2002. Monovalent cations mediate

formation of native tertiary structure of the *Tetrahymena thermophila* ribozyme. Nat. Struct. Biol. 9:928-933.

50. Zhang, Y., J. Zhang, and W. Wang. 2011. Atomistic analysis of pseudoknotted RNA unfolding. J. Am. Chem. Soc. 133:6882-6885.

51. Bian, Y., J. Zhang, J. Wang, J. Wang, and W. Wang. 2015. Free energy landscape and multiple folding pathways of an H-Type RNA pseudoknot. PLoS One 10:e0129089.

52. Xu, X. J., and S. J. Chen. 2012. Kinetic mechanism of conformational switch between bistable RNA hairpins. J. Am. Chem. Soc. 134:12499-12507.

53. Chen, J., and W. Zhang. 2012. Kinetic analysis of the effects of target structures on siRNA efficiency. J. Chem. Phys. 137:225102.

54. Shi, Y. Z., F. H. Wang, Y. Y. Wu, and Z. J. Tan. 2014. A coarse-grained model with implicit salt for RNAs: predicting 3D structure, stability and salt effect. J. Chem. Phys. 141:105102.

55. Pabit, S. A., J. L. Sutton, H. Chen, and L. Pollack. 2013. Role of ion valence in the submillisecond collapse and folding of a small RNA domain. Biochemistry 52:1539-1546.

56. Leipply, D., and D. E. Draper. 2011. Effects of $Mg^{2+}$ on the free energy landscape for folding a purine riboswitch RNA. Biochemistry. 50: 2790-2799.

57. Meisburger, S. P., S. A. Pabit, and L. Pollack. 2015. Determining the locations of ions and water around DNA from X-ray scattering measurements. Biophys. J. 108:2886-2895.

58. Tan, Z. J., and S. J. Chen. 2006. Nucleic acid helix stability: effects of salt concentration, cation valence and size, and chain length. Biophys. J. 90:1175-1190.

59. Tan, Z. J., and S. J. Chen. 2010. Predicting ion binding properties for RNA tertiary structures. Biophys. J. 99:1565-1576.

60. Tan, Z. J., and S. J. Chen. 2007. RNA helix stability in mixed $Na^+/Mg^{2+}$ solution. Biophys. J. 92:3615–3632.

61. Hayes, R. L., J. K. Noel, A. Mandic, P. C. Whitford, K. Y. Sanbonmatsu, U. Mohanty, and J. N. Onuchic. 2015. Generalized manning condensation model captures the RNA ion atmosphere. Phys. Rev. Lett. 114:258105.

62. Hayes, R. L., J. K. Noel, P. C. Whitford, U. Mohanty, K. Y. Sanbonmatsu, and J. N. Onuchic. 2014. Reduced Model Captures $Mg^{2+}$-RNA Interaction Free Energy of Riboswitches. Biophys. J. 106:1508-1519.

63. Hayes, R. L., J. K. Noel, U. Mohanty, P. C. Whitford, S. P. Hennelly, J. N. Onuchic, and K. Y. Sanbonmatsu. 2012. Magnesium fluctuations modulate RNA dynamics in the SAM-I riboswitch. J. Am. Chem. Soc. 134:12043-12053.

64. Hagerman, P. J. 1997. Flexibility of RNA. Annu. Rev. Biophys. Biomol. Struct. 26:139-156.

65. Herschlag, D., B. E. Allred, and S. Gowrishankar. 2015. From static to dynamic: the need for structural ensembles and a predictive model of RNA folding and function. Curr. Opin. Struct. Biol. 30:125-133.

66. Mouzakis, K. D., E. A. Dethoff, M. Tonelli, H. M. Al-Hashimi, and S. E. Butcher. 2015. Dynamic motions of the HIV-1 frameshift site RNA. Biophys. J. 108:644-654.

67. Casiano-Negroni, A., X. Sun, and H. M. Al-Hashimi. 2007. Probing $Na^+$-induced changes in the HIV-1 TAR conformational dynamics using NMR residual dipolar couplings: new insights into the role of counterions and electrostatic interactions in adaptive recognition. Biochemistry. 46:6525-6535.

68. Zacharias, M., and P. J. Hagerman. 1995. Bulge-induced bends in RNA: quantification by transient electric birefringence. J. Mol. Biol. 247:486-500.

69. Chen, H., S. P. Meisburger, S. A. Pabit, J. L. Sutton, W. W. Webb, L. Pollack. 2012. Ionic strength-dependent

persistence lengths of single-stranded RNA and DNA. Proc. Natl. Acad. Sci. U. S. A. 109:799-804.

70. de Pablo, J. J. 2011. Coarse-Grained Simulations of Macromolecules: From DNA to Nanocomposites. Annu. Rev. Phys. Chem. 62:555.

71. Zhou, H. X. 2014. Theoretical frameworks for multiscale modeling and simulation. Curr. Opin. Struct. Biol. 25:67–76.

72. Kikot, I. P., A. V. Savin, E. A. Zubova, M. A. Mazo, E. B. Gusarova, L. I. Manevitch, and A. V. Onufriev. 2011. New coarse-grained DNA model. Biophysics, 56:387-392.

73. Daily, M. D., B. N. Olsen, P. H. Schlesinger, D. S. Ory, and N. A. Baker. 2014. Improved coarse-grained modeling of cholesterol-containing lipid bilayers. J. Chem. Theory Comput. 10:2137-2150.

74. Mustoe, A. M., H. M. Al-Hashimi, and C. L. Brooks. 2014. Coarse grained models reveal essential contributions of topological constraints to the conformational free energy of RNA bulges. J. Phys. Chem. B 118:2615-2627.

75. Wang, F. H., Y. Y. Wu, and Z. J. Tan. 2013. Salt contribution to the flexibility of single-stranded nucleic acid of finite length. Biopolymers. 99:370-381.

76. Huang, S. Y., and X. Q. Zou. 2014. A knowledge-based scoring function for protein-RNA interactions derived from a statistical mechanics-based iterative method. Nucleic Acids Res. 42:e55.

77. Huang, S. Y., and X. Q. Zou. 2011. Statistical mechanics-based energy scoring function for structural model selection in protein structure prediction. Proteins 79:2648–2661.

78. Zhang, Y., H. Zhou, and Z. Ouyang. 2001. Stretching single-stranded DNA: Interplay of electrostatic, base-pairing, and base-pair stacking interactions. Biophys. J. 81:1133-1143.

79. Xia, T., J. SantaLucia, M. E. Burkand, R. Kierzek, S. J. Schroeder, X. Jiao, C. Cox, and D. H. Turner. 1998. Thermodynamic parameters for an expanded nearest-neighbor model for formation of RNA duplexes with Watson-Crick base pairs. Biochemistry 37:14719-14735.

80. Mathews, D. H., J. Sabina, M. Zuker, and D. H. Turner. 1999. Expended sequence dependence of thermodynamic parameters improves prediction of RNA secondary structure. J. Mol. Biol. 288:911-940.

81. Thomas, D. G., J. Chun, Z. Chen, G. Wei, and N. A. Baker. 2013. Parameterization of a geometric flow implicit solvation model. J. Comput. Chem. 34:687-695.

82. Ren, P., J. Chun, D. G. Thomas, M. Schnieders, M. Marucho, J. Zhang, and N. A. Baker. 2012. Biomolecular electrostatics and solvation: a computational perspective." Q. Rev. Biophys. 45:427-491.

83. Manning, G. S. 1978. The molecular theory of polyelectrolyte solutions with applications to the electrostatic properties of polynucleotides. Q. Rev. Biophys. 11:179-246.

84. Tan, Z. J., and S. J. Chen. 2005. Electrostatic correlations and fluctuations for ion binding to a finite length polyelectrolyte. J. Chem. Phys. 122:44903.

85. Tan, Z. J., and S. J. Chen. 2011. Salt contribution to RNA tertiary structure folding stability. Biophys. J. 101:176-187.

86. Walter, A. E., D. H. Turner, J. Kim, M. H. Lyttle, P. Müller, D. H. Mathews, and M. Zuker. 1994. Coaxial stacking of helixes enhances binding of oligoribonucleotides and improves predictions of RNA folding. Proc. Natl. Acad. Sci. 91:9218–9222.

87. Walter, A. E., and D. H. Turner. 1994. Sequence dependence of stability for coaxial stacking of RNA helixes with Watson-Crick base paired interface. Biochemistry 33:12715–12719.

88. Kirkpatrick, S., C. D. Gelatt, and M. P. Vecchi. 1983. Optimization by simulated annealing. Science. 220:671-680.
89. Parisien, M., J. A. Cruz, E. Westhof, and F. Major. 2009. New metrics for comparing and assessing discrepancies between RNA 3D structures and models. RNA 15:1875-1885.
90. Zhang, Y., and J. Skolnick. 2004. Scoring function for automated assessment of protein structure template quality. Proteins. 57:702-710.
91. Stallings, S. C., and P. B. Moore. 1997. The structure of an essential splicing element: stem loop IIa from yeast U2 snRNA. Structure. 5:1173-1185.
92. Tan, Z. J., and S. J. Chen. 2008. Electrostatic free energy landscapes for DNA helix bending. Biophys. J. 94:3137-3149.
93. Wu, Y. Y., L. Bao, X. Zhang, and Z. J. Tan. 2015. Flexibility of short DNA helices with finite-length effect: from base pairs to tens of base pairs. J. Chem. Phys. 142:125103.
94. Kuznetsov, S. V., C. C. Ren, S. A. Woodson, and A. Ansari. 2008. Loop dependence of the stability and dynamics of nucleic acid hairpins. Nucleic Acids Res. 36:1098-1112.
95. Sehdev, P., G. Crews, and A. M. Soto. 2012. Effect of helix stability on the formation of loop-loop complexes. Biochemistry. 51:9612-9623.
96. Tan, Z. J., and S. J. Chen. 2008. Salt dependence of nucleic acid hairpin stability. Biophys. J. 96:738-752.
97. Wu, Y. Y., Z. L. Zhang, J. S. Zhang, X. L. Zhu, and Z. J. Tan. 2015. Multivalent ion-mediated nucleic acid helix-helix interactions: RNA versus DNA. Nucleic Acids Res. 43:6156-6165.
98. Nixon, P. L., C. A. Theimer, and D. P. Giedroc. 1999. Thermodynamics of stabilization of RNA pseudoknots by cobalt[III] hexamine. Biopolymers. 50:443–458.
99. Nixon, P. L., and D. P. Giedroc. 1998. Equilibrium unfolding (folding) pathway of a model H-type pseudoknotted RNA: the role of magnesium ions in stability. Biochemistry. 37:16116–16129.

**FIGURES AND TABLES**

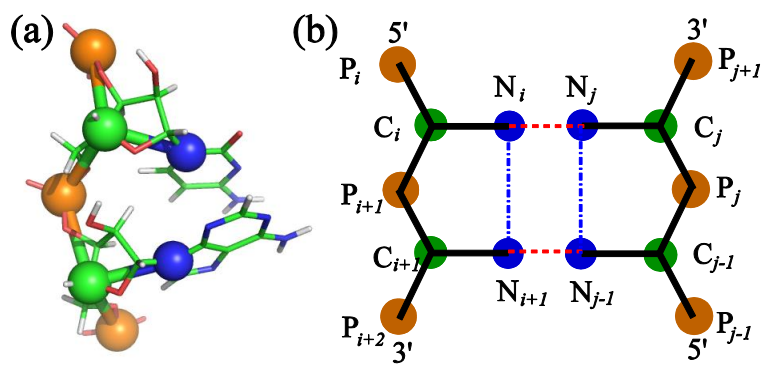

**FIGURE 1.** (a) Our coarse-grained representation for one fragment of an RNA superposed on an all-atom representation. Namely, three beads are located at the atoms of phosphate (P), C4' (C), and N1 for pyrimidine or N9 for purine (N), respectively. The structure is shown with the PyMol (http://www.pymol.org). (b) The schematic representation for base-pairing (dashed line) and base-stacking (dash-dotted line).

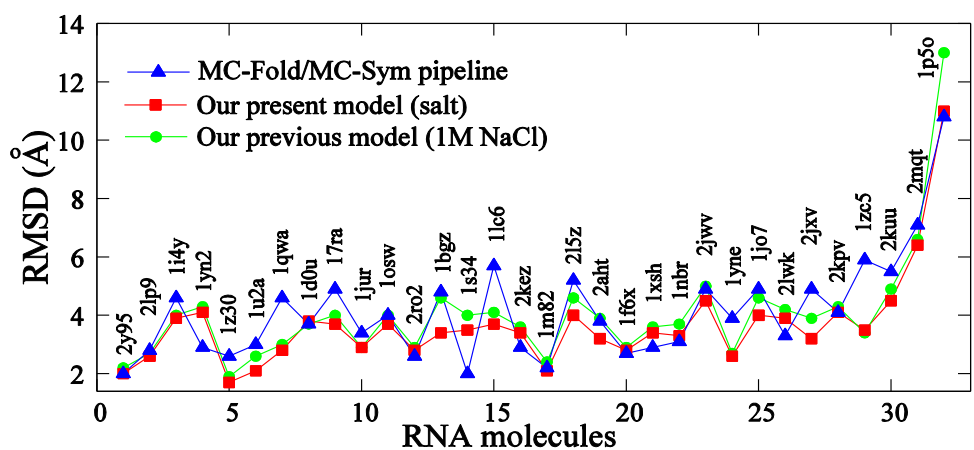

**FIGURE 2.** (a) Comparison of the RMSDs of predicted RNA 3D structures between the present/previous version of our model and the MC-Fold/MC-Sym pipeline. For each of the 32 tested RNAs, we use the MC-Fold/MC-Sym pipeline online tool (http://www.major.iric.ca/MC-Fold/) (30) to test the accuracy of MC-Fold/MC-Sym and calculate the RMSD for the top 1 predicted structure over C4' atom in the backbone. The RMSDs of structures predicted by the present model at the experimental ion conditions (see in Table I) and by its previous version (54) at 1M NaCl are calculated over C beads from the corresponding C4' atoms in the native structures.

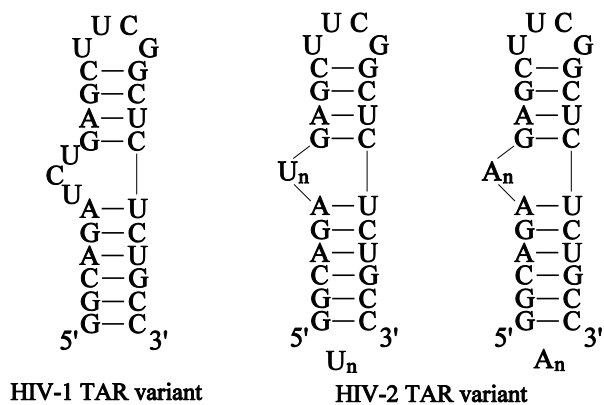

**FIGURE 3.** The sequences and the secondary structures predicted by the present model for HIV-1 TAR variant with a 3-nt bulge and HIV-2 TAR variants with different length of polyU (polyA) bulges, which are used to study the flexibility of RNAs.

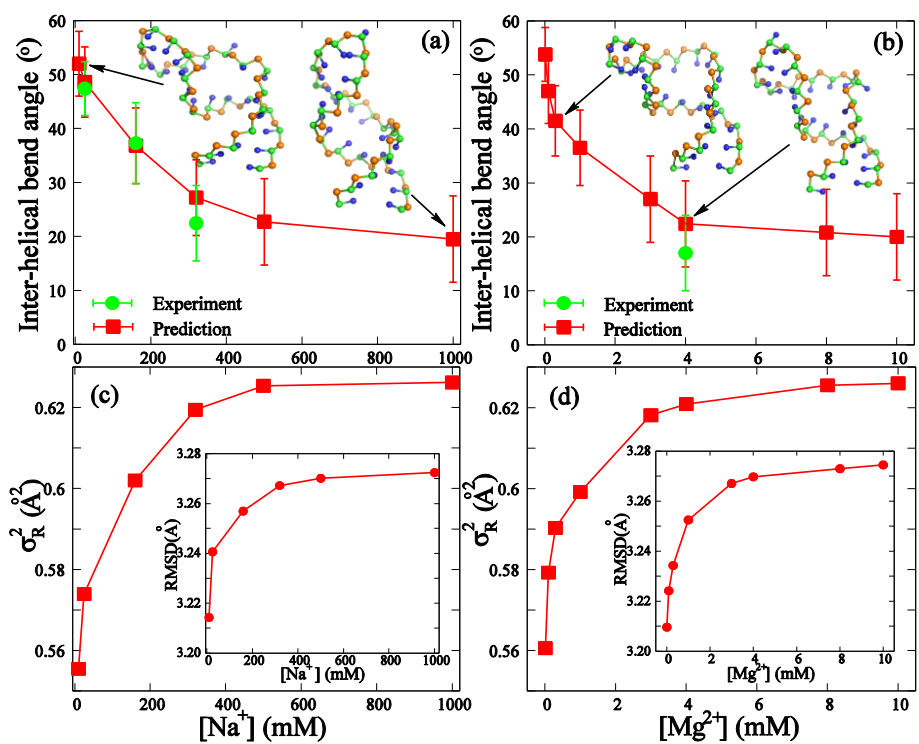

**FIGURE 4**. (a, b) The experimental (Ref. 67) and predicted inter-helical bend angle as functions of [$Na^+$] (a) and [$Mg^{2+}$] (b) for HIV-1 TAR variant (see Fig. 3). The corresponding typical 3D structures predicted by the present model are shown with the PyMol (http://www.pymol.org). (c, d) The variances $\sigma_R^2$ of radius of gyration and the RMSDs from time-averaged reference structures as functions of [$Na^+$] (c) and [$Mg^{2+}$] (d) for HIV-1 TAR variant (see Fig. 3).

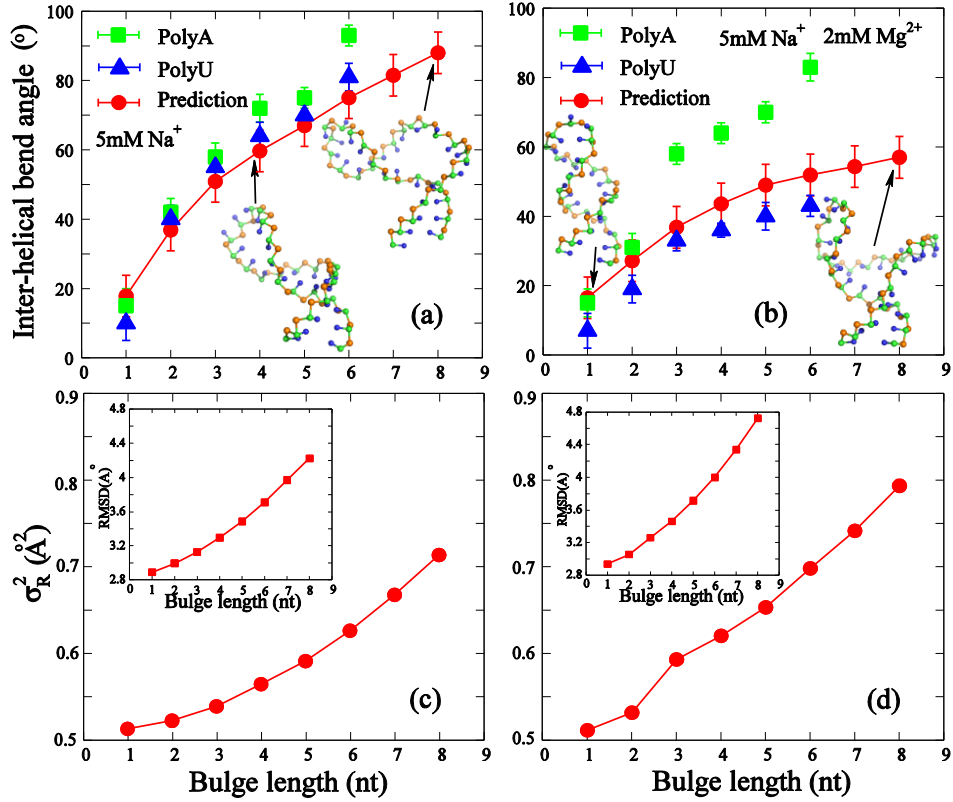

**FIGURE 5.** (a, b) The experimental (Ref. 68) and predicted inter-helical bend angles as functions of bulge length at 5mM NaPO$_4$ without (a) or with (b) 2mM Mg$^{2+}$ for HIV-2 TAR variant (see Fig. 3). The corresponding typical 3D structures predicted by the present model are shown with the PyMol (http://www.pymol.org). (c, d) The variances $\sigma_R^2$ of radius of gyration and the RMSDs from time-averaged reference structures as functions of bulge length at 5mM NaPO$_4$ without (c) or with 2mM Mg$^{2+}$ (d) for HIV-2 TAR variant (see Fig. 3).

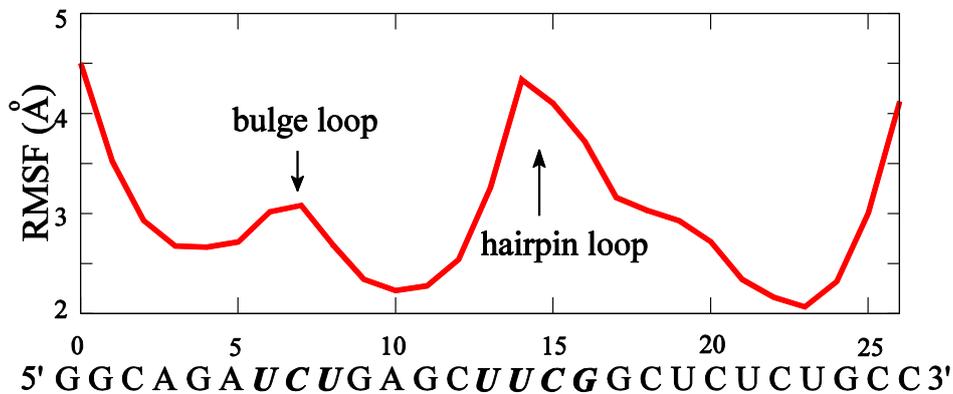

**FIGURE 6.** The RMSF for C-beads along HIV-1 TAR variant (see Fig. 3) in 1M NaCl solution. The sequences of bulge and hairpin loops are in italics.

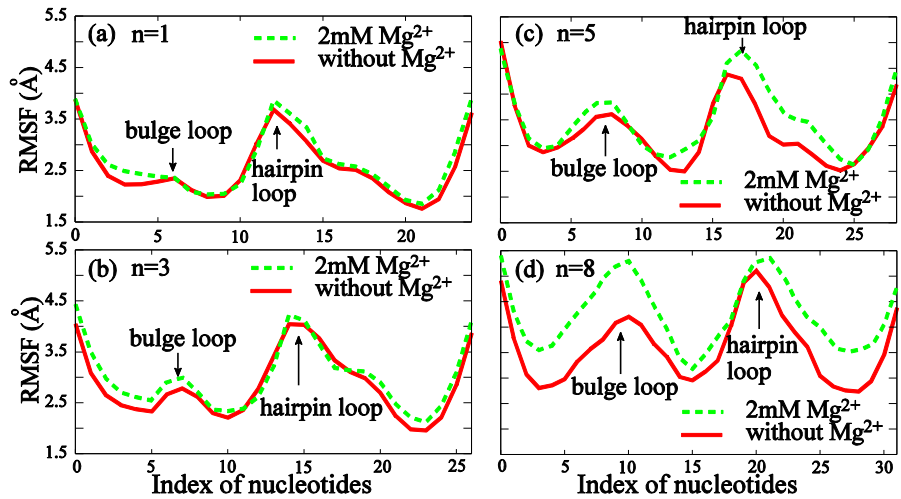

**FIGURE 7.** The RMSF for C-beads along HIV-2 TAR variant (see Fig. 3) with different bulge length at 5mM NaPO$_4$ with/without 2mM Mg$^{2+}$. The length $n$ of bulge loop is 1 (a), 3 (b), 5 (c) and 8 (d), respectively.

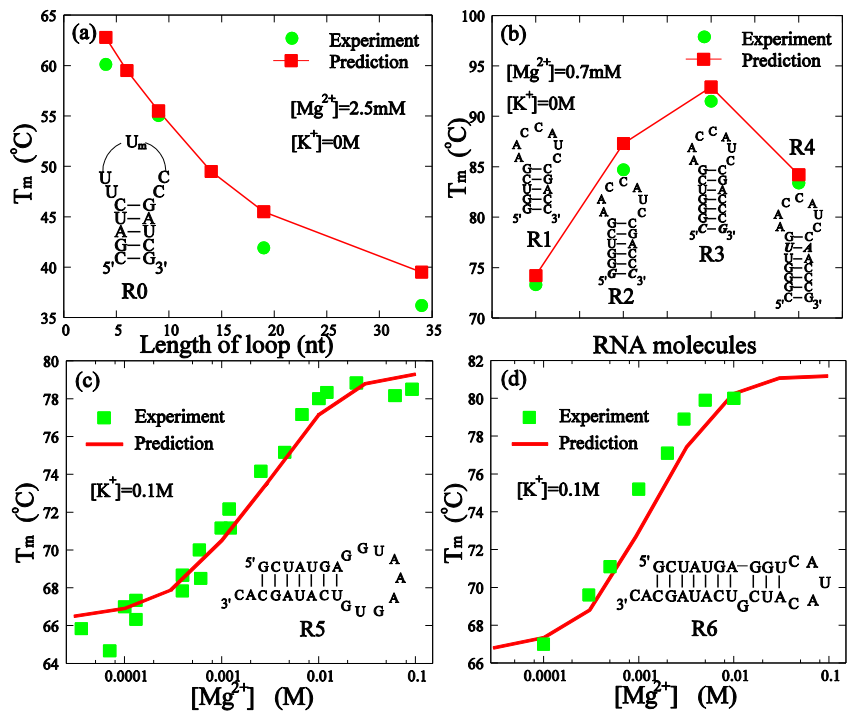

**FIGURE 8.** (a) The experimental (Ref. 94) and predicted melting temperature as functions of length of hairpin loop for RNA hairpin R0 at 2.5mM [Mg$^{2+}$]. (b) The experimental (Ref. 95) and predicted melting temperatures of four RNA hairpins with different stems at 0.7mM [Mg$^{2+}$]. (c, d) The experimental (Ref. 98 and Ref. 99) and predicted melting temperature $T_m$'s as functions of [Mg$^{2+}$] for two RNA hairpins R5 (c) and R6 (d) in the presence of 0.1 M [K$^+$]. The sequences and the secondary structures predicted by the present model are also shown in (a), (b), (c) and (d).

Table I. The 32 RNA molecules for 3D structure prediction in this work.

| RNAs | PDB[a] | Length (nt) | Type[d] | [1+/2+][e] (mM) | RMSD$_{pred.}$[f] (Å) (mean/minimum) | RMSD$^0_{pred.}$[g] (Å) | RMSD$_{MC-Sym}$[h] (Å) |
|---|---|---|---|---|---|---|---|
| 1 | 2y95[b] | 14 | H | 100/0 | 2.0/1.0 | 2.2 | 2 |
| 2 | 2lp9[b] | 16 | B | 65/0 | 2.6/1.1 | 2.7 | 2.8 |
| 3 | 1j4y[b] | 17 | H | 20/0 | 3.9/1.9 | 4 | 4.6 |
| 4 | 1yn2[b] | 17 | H | 55/40 | 4.1/2.3 | 4.3 | 2.9 |
| 5 | 1z30[b] | 18 | H | 50/0 | 1.7/0.9 | 1.9 | 2.6 |
| 6 | 1u2a[c] | 20 | H | 75/0 | 2.1/1.1 | 2.6 | 3 |
| 7 | 1qwa[b] | 21 | B | 5/0 | 2.8/1.5 | 3 | 4.6 |
| 8 | 1d0u[b] | 21 | B | 50/0 | 3.8/1.5 | 3.7 | 3.7 |
| 9 | 17ra[c] | 21 | B | 20/0 | 3.7/1.3 | 4 | 4.9 |
| 10 | 1jur[b] | 22 | B | 100/0 | 2.9/1.5 | 2.9 | 3.4 |
| 11 | 1osw[b] | 22 | I | 25/0 | 3.7/1.5 | 4 | 4 |
| 12 | 2ro2[b] | 23 | H | 12.4/0 | 2.8/1.4 | 2.9 | 2.6 |
| 13 | 1bgz[c] | 23 | B&I | 20/0 | 3.4/2.1 | 4.6 | 4.8 |
| 14 | 1s34[b] | 23 | B | 35/0 | 3.5/1.7 | 4 | 2 |
| 15 | 1lc6[b] | 24 | I | 50/0 | 3.7/2.0 | 4.1 | 5.7 |
| 16 | 2kez[b] | 24 | I | 50/0 | 3.4/1.5 | 3.6 | 2.9 |
| 17 | 1m82[b] | 25 | B | 20/0 | 2.1/1.2 | 2.4 | 2.2 |
| 18 | 2l5z[b] | 26 | I | 50/5 | 4.0/2.6 | 4.6 | 5.2 |
| 19 | 2aht[b] | 27 | B | 100/6 | 3.2/1.3 | 3.9 | 3.8 |
| 20 | 1f6x[b] | 27 | B | 100/0 | 2.8/1.5 | 2.9 | 2.7 |
| 21 | 1xsh[b] | 27 | B | 100/0 | 3.4/1.7 | 3.6 | 2.9 |
| 22 | 1nbr[b] | 29 | B | 20/0 | 3.3/2.1 | 3.7 | 3.1 |
| 23 | 2jwv[b] | 29 | I | 50/0 | 4.5/2.0 | 5 | 4.9 |
| 24 | 1yne[b] | 31 | B | 10/0 | 2.6/1.3 | 2.7 | 3.9 |
| 25 | 1jo7[b] | 31 | B&I | 10/0 | 4.0/2.3 | 4.6 | 4.9 |
| 26 | 2lwk[b] | 32 | B&I | 50/0 | 3.9/1.7 | 4.2 | 3.3 |
| 27 | 2jxv[b] | 33 | I | 20/0 | 3.2/1.9 | 3.9 | 4.9 |
| 28 | 2kpv[b] | 34 | B&I | 20/0 | 4.1/1.9 | 4.3 | 4.1 |
| 29 | 1zc5[b] | 41 | B | 10/0 | 3.5/1.8 | 3.4 | 5.9 |
| 30 | 2kuu[b] | 48 | B | 20/0 | 4.5/2.3 | 5.1 | 5.5 |
| 31 | 2mqt[b] | 68 | B&I | 10/0 | 6.4/3.8 | 6.8 | 7.1 |
| 32 | 1p5o[b] | 77 | B&I | 100/5 | 11.0/8.7 | 13 | 10.8 |

[a] The 3D structures of these RNA hairpins have been determined by NMR method at certain ion conditions. [b, c] The hairpins are experimentally determined after[b] and before[c] the year of 2000, respectively. [d] H, B, I, B&I represent RNA hairpins without bulge/internal loops (H), RNA hairpins with bulges (B), RNA hairpins with internal loops (I), and RNA hairpins with bulge and internal loops (B&I), respectively. [e] The salt conditions of solutions in which RNA structure have been experimentally determined. [f] The mean/minimum RMSDs are calculated over C beads of the structures predicted by the present model from the corresponding atoms C4' of the native structures. [g] The mean RMSDs are calculated over C beads of structures predicted by our previous model at 1M NaCl from the corresponding atoms C4' of the native structures. [h] The RMSD is calculated over the C4' atoms of the top 1 structure for each RNA predicted by the MC-Fold/MC-Sym pipeline (http://www.major.iric.ca/MC-Fold/) (30) from the native structure.